\documentclass[12pt]{article}
\usepackage{epsfig}
\usepackage{amsmath}
\def\Journal#1#2#3#4{{#1}{\bf #2}, #3 (#4)}

\def\NPB{{\em Nucl. Phys.~}{\bf B}}
\def\PLB{{\em Phys. Lett.~}{\bf  B}}

\def\PRD{{\em Phys. Rev.~}{\bf D}}
\def\ZPC{{\em Z. Phys.~}{\bf C}}
\def\ZP{\em Z. Phys.~}
\def\JHEP{\em J. High Energy Phys.~}
\def\CPC{\em Computer Phys. Commun.~}
\def\PRP{\em Phys. Rep.~}
\def\PRV{\em Phys. Rev.~}
\def\EPJC{{\em Eur. Phys. J.~}{\bf C}}
\newcommand{\aem}{\alpha_{\mathrm{em}}}
\renewcommand{\d}{\mathrm{d}}
\newcommand{\X}{\mathbf{X}}
\newcommand{\e}{\mathrm{e}}

\newcommand{\g}{\mathrm{g}}
\newcommand{\p}{\mathrm{p}}
\newcommand{\q}{\mathrm{q}}

\newcommand{\qbar}{\mathrm{\overline{q}}}

\newcommand{\kT}{k_{\perp}}
\newcommand{\pT}{p_{\perp}}
\newcommand{\shat}{\hat{s}}

\newcommand{\gast}{\gamma^*}
\newcommand{\gtrsim}{\raisebox{-0.8mm}%
{\hspace{1mm}$\stackrel{>}{\sim}$\hspace{1mm}}}

\newlength{\abstwidth}
\begin{document}
 
\sloppy
 
\pagestyle{empty}

\begin{flushright}
LU TP 99--12 \\
hep-ph/9906407\\
June 1999
\end{flushright}
 
\vspace{\fill}

\begin{center}  \begin{Large} \begin{bf}
Jet Production by Real and Virtual Photons%
\footnote{To appear in the Proceedings of the DESY Workshop on %
Monte Carlo Generators for HERA Physics}\\[10mm]
  \end{bf}  \end{Large}
  \vspace*{5mm}
  \begin{large}
C. Friberg and T. Sj\"ostrand\\[2mm]
  \end{large}
 Department of Theoretical Physics, Lund University,\\ 
     Helgonav\"agen 5, S-223 62 Lund, Sweden\\ 
     christer@thep.lu.se, torbjorn@thep.lu.se
\end{center}

\vspace{\fill}

\begin{center}
{\bf Abstract}\\[2ex]
\begin{minipage}{\abstwidth}
The production of jets is studied in collisions of virtual photons, 
specifically for applications at HERA. 
Photon flux factors are convoluted with matrix elements involving
either direct or resolved photons and, for the latter, with parton
distributions of the photon. Special emphasis is put on the
range of uncertainty in the modeling of the resolved component.
The resulting model is compared with existing data and further 
tests are proposed.
\end{minipage}
\end{center}

\vspace{\fill}

\clearpage
\pagestyle{plain}
\setcounter{page}{1} 

\section{Introduction}

The photon is a complicated object to describe. In the DIS region, i.e. 
when it is very virtual, it can be considered as devoid of any internal
structure, at least to first approximation. In the other extreme, the 
total cross section for real photons is dominated by the resolved 
component of the wave function, where the photon has fluctuated 
into a $\q\qbar$ state. The nature of this resolved component is
still not well understood, especially not the way in which it dies 
out with increasing photon virtuality. This dampening is likely not 
to be a simple function of virtuality, but to depend on the physics
observable being studied, i.e. on the combination of subprocesses
singled out.  

Since our current understanding of QCD does not allow complete 
predictability, one sensible approach is to base ourselves on
QCD-motivated models, where a plausible range of uncertainty can be 
explored. Hopefully comparisons with data may then help constrain
the correct behaviour. The ultimate goal therefore clearly is to
have a testable model for all aspects of the physics of $\gast\p$ 
(and $\gast\gast$) collisions. As a stepping stone towards 
constructing such a framework, in this paper we explore the physics 
associated with the production of `high-$\pT$' jets in the collision. 
That is, we here avoid the processes that only produce activity along 
the $\gast\p$ collision axis. For resolved photons this
corresponds to the `soft' or `low-$\pT$' events of the hadronic 
physics analogy, for direct ones to the lowest-order DIS process
$\gast\q \to \q$. 

The processes that we will study here instead can be exemplified by
$\gast \g \to \q\qbar$ (direct) and 
$\g\g \to \q\qbar$ (resolved), where the gluons come from the 
parton content of a resolved virtual photon or from the proton.
Note that these are multi-scale processes, at least involving the
virtuality $Q^2$ of the photon and the $\pT^2$ of 
the hard subprocess.
For a resolved photon, the relative transverse momentum $\kT$ of 
the initial $\gast \to \q\qbar$ branching provides a further scale,
at least in our framework. This plethora of scales clearly is a 
challenge to any model builder, but in principle it also offers 
the opportunity to explore QCD in a more differential fashion than 
is normally possible. 

At large photon virtualities, a possible strategy would be to express
the cross sections entirely in terms of processes involving the photon
directly, i.e. to include branchings such as $\gast \to \q'\qbar'$ and
$\q' \to \q'\g$ in the Feynman graphs calculated to describe the
process, so that e.g. the $\gast\p$ process $\g\g \to \q\qbar$ is
calculated as $\gast\g \to \q'\qbar'\q\qbar$. With decreasing
virtuality of the photon, such a fixed-order approach is increasingly
deficient by its lack of the large logarithmic corrections generated
by collinear and soft gluon emission, however.  Furthermore, almost
real photons allow long-lived $\gast \to \q\qbar$ fluctuations, that
then take on the properties of non-perturbative hadronic states,
specifically of vector mesons such as the $\rho^0$.  It is therefore
that an effective description in terms of parton distributions becomes
necessary. Hence the resolved component of the photon, as opposed to
the direct one.

That such a subdivision is more than a technical construct is
excellently illustrated by the $x_{\gamma}^{\mathrm{obs}}$ plots from
HERA \cite{xgobs}. This variable sums up the fraction of the original
photon lightcone momentum carried by the two highest-$E_{\perp}$
jets. A clear two-component structure is visible.  The peak close to
$x_{\gamma}^{\mathrm{obs}} = 1$ can be viewed as a smeared footprint
of the direct photon, with all the energy partaking in the hard
interaction, while the broad spectrum at lower
$x_{\gamma}^{\mathrm{obs}}$ is consistent with the resolved photon,
where much of the energy remains in a beam jet. The distinction
between the two is not unique when higher-order effects are included,
but it is always possible to make a functional separation that avoids
double-counting or gaps.

The resolved photon can be further subdivided into low-virtuality
fluctuations, which then are of a nonperturbative character and can be
represented by a set of vector mesons, and high-virtuality ones that
are describable by perturbative $\gast \to \q\qbar$ branchings.  The
former is called the VMD (vector meson dominance) component and the
latter the anomalous one. The parton distributions of the VMD
component are unknown from first principles, and thus have to be based
on reasonable ans\"atze, while the anomalous ones are perturbatively
predictable. This separation is more ambiguous and less well tested
than the one between direct and resolved photons. In principle,
studies on the structure of the beam remnant, e.g. its $\pT$
distribution, should show characteristic patterns. Unfortunately, the
naively expected differences are smeared by higher-order QCD
corrections (especially initial-state radiation), by the possibility
of multiple parton--parton interactions, by hadronization effects, and
so on. (Experimentally, gaps in the detector acceptance, e.g. for the
beam pipe, is a further major worry.) Many of these areas offer
interesting challenges in their own right; e.g. the way in which
multiple interactions die out with virtuality, both that of the photon
itself and that of the $\q\qbar$ pair it fluctuates to.  Models for
one aspect at the time are therefore likely to be inadequate. Instead
we here attempt a combined description of all the relevant physics
topics.

The traditional tool for handling such complex issues is the Monte Carlo
approach. Our starting point is the model for real photons 
\cite{sasevt} and the parton distribution parameterizations of real
and virtual photons \cite{saspdf} already present in the {\sc Pythia}
\cite {pythia} generator. Several further additions and modifications
have been made to model virtual photons, as will be described in the
following. Other generators with an overlapping scope include, 
among others, HERWIG \cite{herwig}, LDC \cite{ldc}, LEPTO \cite{lepto},
PHOJET \cite{phojet} and RAPGAP \cite{rapgap}. The details of the 
approaches are different, 
however, so this gives healthy possibilities to compare and learn.
Another alternative is provided by matrix-element
calculations \cite{nlome}, that do not provide the same complete 
overview but can offer superior descriptions for some purposes.

The plan of this paper is the following. In section 2 the model is
described, with special emphasis on those aspects that are new compared
with the corresponding description for real photons. Thereafter, in
section 3, comparisons are shown with some sets of data from HERA, 
and this is used to constrain partly the freedom in the model. Finally, 
section 4 contains a summary and outlook.

\section{The Model}
\label{model}

The electromagnetic field surrounding a moving electron can be 
viewed as a flux of photons. Weizs\"acker \cite{weiz} and Williams 
\cite{will} calculated the spectrum of these photons, neglecting 
the photon virtualities and terms involving the longitudinal polarization 
of photons. This approximation is well-known \cite{klaWW} to be a good 
approximation when the scattered lepton is tagged at small scattering 
angles.

In the equivalent photon approximation \cite{EPA}, the cross sections 
for the process $\e\p\rightarrow\e\X$, 
where $\X$ is an arbitrary final state, can then be written as the 
convolution
$\d\sigma(\e\p\rightarrow\e\X)=\int \d \omega \; N(\omega) 
\;\d\sigma(\gamma\p\rightarrow\X)$
where $\omega$ is the energy of the emitted photon. In this approximation, 
the distribution in photon frequencies $N(\omega)\d \omega$ is 
obtained by integrating over the photon virtuality $Q^2$. The maximum 
value $Q^2_{\mathrm{max}}$ is usually given by experimental conditions like 
anti-tagging, i.e. that the scattered lepton is not detected if its 
scattering angle is too small.

A better approximation, and the one used in our approach, 
is to keep the $Q^2$ dependence in the photon flux 
$f(y,Q^2)$ (with $y \approx \omega/\omega_{\mathrm{max}}$, see below) 
and in the subprocess cross sections involving the virtual 
photon, $\gast\p\rightarrow\X$,
and to sum over the transverse and longitudinal photon polarizations. 
We then arrive with
\begin{equation}
\d\sigma(\e\p\rightarrow\e\X)=\sum_{\xi=\mathrm{T,L}}
\iint \d y \, \d Q^2 
\;f_{\gamma/\e}^{\xi}(y,Q^2) 
\;\d\sigma(\gast_{\xi}\p\rightarrow\X)\;.
\label{PYHERAEPA}
\end{equation}
This factorized ansatz is perfectly general, so long 
as azimuthal distributions in the final state are not studied in detail.

When $Q^2/W^2$ is small, one can derive~\cite{gammaflux,EPA,GS}
\begin{eqnarray}
f_{\gamma/\e}^{\mathrm{T}}(y,Q^2) =  \frac{\aem}{2\pi} 
\left( \frac{(1+(1-y)^2}{y} \frac{1}{Q^2}-\frac{2m_{\e}^2y}{Q^4} \right) 
~,~~~~~
f_{\gamma/\e}^{\mathrm{L}}(y,Q^2) =  \frac{\aem}{2\pi} 
\frac{2(1-y)}{y} \frac{1}{Q^2}\;. 
\label{LLogflux}
\end{eqnarray}
The $y$ variable is defined as the lightcone fraction the photon takes 
of the incoming lepton momentum. In the $\e\p$ kinematics, the $y$ 
definition gives that
\begin{equation}
y = \frac{q P}{k P} ~,~~~~~ W^2 = y s - Q^2 ~. 
\end{equation}
(Here and in the following formulae we have omitted the lepton and hadron 
mass terms when it is not of importance for the argumentation.) The lepton 
scattering angle $\theta$ is related to $Q^2$, where the kinematical limits 
on $Q^2$ are, unless experimental conditions reduce the $\theta$ range, 
$Q^2_{\mathrm{min}} \approx \frac{y^2}{1 - y} m_{\e}^2$ and
$Q^2_{\mathrm{max}} \approx (1 - y) s$. 

In summary, we will allow the possibility of experimental cuts in the
$y$, $Q^2$, $\theta$ and $W^2$ variables. Within the 
allowed region, the phase space is Monte Carlo sampled according to
$(\d Q^2/Q^2) \, (\d y / y) \, \d \varphi$, with the 
remaining flux factor combined with the cross section factors to give 
the event weight used for eventual acceptance or rejection.

The hard-scattering processes are classified according to whether 
the photon is resolved or not. For the direct processes,
QCD Compton $\gast \q \to \g \q$
and boson--gluon fusion $\gast \g \to \q \qbar$, 
both transverse and longitudinal photons are considered. The matrix 
elements are given elsewhere~\cite{siggap,paper}. Remember that the 
cross section for a longitudinal photon vanishes as $Q^2$ in the limit 
$Q^2 \to 0$. 

For a resolved photon, there are six basic QCD cross sections, 
$\q\q' \to \q\q'$, $\q\qbar \to \q'\qbar'$, $\q\qbar \to \g\g$,
$\q\g \to \q\g$, $\g\g \to \g\g$ and $\g\g \to \q\qbar$.
The photon virtuality scale is included in the 
arguments of the parton distribution but, in the spirit of the parton 
model, the virtuality of the parton inside the photon is not included 
in the matrix elements. Neither is the possibility of the partons being
polarized. The same subprocess cross sections as those known from $\p\p$ 
physics \cite{sigpp} can therefore be used for resolved $\gast \p$ 
processes. A convolution with parton distributions is then necessary. 

One major element of model dependence enters via the choice of parton
distributions for a resolved virtual photon. These distributions
contain a hadronic component that is not perturbatively calculable. It
is therefore necessary to parameterize the solution with input from
experimental data, which mainly is available for (almost) real
photons. In the following we will use the SaS distributions
\cite{saspdf}, which are the ones best suited for our
formalism. Another set of distributions is provided by GRS
\cite{grspdf}, while a simpler recipe for suppression factors relative
to real photons has been proposed by DG \cite{dgpdf}.

The SaS distributions for a real photon can be written as
\begin{equation}
f_a^{\gamma}(x,\mu^2) =
\sum_V \frac{4\pi\aem}{f_V^2} f_a^{\gamma,V}(x,\mu^2; Q_0^2)
+ \frac{\aem}{2\pi} \, \sum_{\q} 2 e_{\q}^2 \,
\int_{Q_0^2}^{\mu^2} \frac{{\d} k^2}{k^2} \,
f_a^{\gamma,\q\qbar}(x,\mu^2;k^2) ~.
\label{decomp}
\end{equation}
Here the sum is over a set of vector mesons
$V = \rho^0, \omega, \phi, \mathrm{J}/\psi$ according to a 
vector-meson-dominance ansatz for low-virtuality fluctuations of
the photon, with experimentally determined couplings $4\pi\aem/f_V^2$.
The higher-virtuality, perturbative, fluctuations are 
represented by an integral over the virtuality $k^2$ and a sum over
quark species. We will refer to the first part as the VMD one and 
the second as the anomalous one.  

From the above ansatz, the extension to a virtual photon is given by
the introduction of a dipole dampening factor for each component,
\begin{eqnarray}
f_a^{\gast}(x,\mu^2,Q^2)
& = & \sum_V \frac{4\pi\aem}{f_V^2} \left(
\frac{m_V^2}{m_V^2 + Q^2} \right)^2 \,
f_a^{\gamma,V}(x,\mu^2;\tilde{Q}_0^2)
\nonumber \\
& + & 
\frac{\aem}{2\pi} \, \sum_{\q} 2 e_{\q}^2 \,
\int_{Q_0^2}^{\mu^2} \frac{{\d} k^2}{k^2} \, \left(
\frac{k^2}{k^2 + Q^2} \right)^2 \, f_a^{\gamma,\q\qbar}(x,\mu^2;k^2)
~.
\label{decompvirt}
\end{eqnarray}
Thus, with increasing $Q^2$, the VMD components die away faster than
the anomalous ones, and within the latter the low-$k^2$ ones faster
than the high-$k^2$ ones. As a technical trick, the handling of the 
$k^2$ integral is made more tractable by replacing the dipole factor 
by a $k^2$-independent multiplicative factor and an increased lower
limit of the integral, in such a way that both the momentum sum
and the average evolution range is unchanged. Finally, correction
factors are introduced to ensure that $f_a^{\gast}(x,\mu^2,Q^2) \to 0$
for $\mu^2 \to Q^2$: in the region $Q^2 > \mu^2$ a fixed-order
perturbative description is more appropriate than the leading-log
description in terms of a resolved photon.  

Since the probed real photon is purely transverse, the above ansatz
does not address the issue of parton distributions of the longitudinal
virtual photons. One could imagine an ansatz based on longitudinally
polarized vector mesons, and branchings $\gast_{\mathrm{L}} \to
\q\qbar$, but currently no parameterization exists along these
lines. We will therefore content ourselves by exploring a simple 
alternative based on applying a simple
multiplicative factor $R$ to the results obtained for a resolved
transverse photon. As usual, processes involving longitudinal photons
should vanish in the limit $Q^2 \to 0$. To study two extremes, the 
region with a linear rise in $Q^2$ is defined either by $Q^2<\mu^2$
or by $Q^2<m_{\rho}^2$, where the former represents the perturbative
and the latter some non--perturbative scale. Also the high-$Q^2$ limit 
is not well constrained; we will compare two different alternatives, 
one with an asymptotic fall-off like $1/Q^2$ and another which 
approaches a constant ratio, both with respect to the transverse resolved 
photon. (Since we put $f_a^{\gast}(x,\mu^2,Q^2) = 0$ for $Q^2>\mu^2$, 
the $R$ value will actually not be used for large $Q^2$, so the 
choice is not so crucial.) We therefore study the alternative 
ans\"atze
\begin{eqnarray}
R_1(y,Q^2,\mu^2) & = & 1 + a \frac{4 \mu^2 Q^2}{(\mu^2 + Q^2)^2}
\frac{f_{\gamma/l}^{\mathrm{L}}(y,Q^2)}
{f_{\gamma/l}^{\mathrm{T}}(y,Q^2)}~,\label{Rfact1}\\
R_2(y,Q^2,\mu^2) & = & 1 + a \frac{4 Q^2}{(\mu^2 + Q^2)}
\frac{f_{\gamma/l}^{\mathrm{L}}(y,Q^2)}
{f_{\gamma/l}^{\mathrm{T}}(y,Q^2)}~,\label{Rfact2}\\
R_3(y,Q^2,\mu^2) & = & 1 + a \frac{4 Q^2}{(m_{\rho}^2 + Q^2)}
\frac{f_{\gamma/l}^{\mathrm{L}}(y,Q^2)}
{f_{\gamma/l}^{\mathrm{T}}(y,Q^2)}\label{Rfact3}
\end{eqnarray}
with $a=1$ as main contrast to the default $a=0$. The $y$ dependence 
compensates for the difference in photon flux between transverse
and longitudinal photons. 

Another ambiguity is the choice of $\mu^2$ scale in parton distributions.
Based on various considerations, we compare six different alternatives: 
\begin{eqnarray}
\mu_1^2 & = & \pT^2 ~,\label{mu1}\\
\mu_2^2 & = & \pT^2 \, \frac{\shat+ x Q^2}{\shat}  ~,
\label{mu2}\\
\mu_3^2 & = & \pT^2 \, \frac{\shat+ Q^2}{\shat} ~,\label{mu3}\\
\mu_4^2 & = & \pT^2 + \frac{Q^2}{2} ~,\label{mu4}\\
\mu_5^2 & = & \pT^2 + Q^2  ~,\label{mu5}\\
\mu_6^2 & = & 2 \mu_3^2~.\label{mu6}
\end{eqnarray}  
Only the fifth alternative ensures $f_a^{\gast}(x,\mu^2,Q^2) > 0$ for
arbitrarily large $Q^2$; in all other alternatives the resolved
contribution (at fixed $\pT$) vanish above some $Q^2$ scale. The last 
alternative exploits the well-known freedom of including some multiplicative
factor in any (leading-order) scale choice.
When nothing is mentioned explicitly below, the choice $\mu_3^2$ is used.

The issues discussed above are the main ones that distinguish the
description of processes involving virtual photons from those 
induced by real photons or by hadrons in general. In common is 
the need to consider the buildup of more complicated partonic
configurations from the lowest-order `skeletons' defined above, 
(\textit{i}) by parton showers, (\textit{ii}) by multiple 
parton--parton interactions and beam remnants, where applicable,
and (\textit{iii}) by the subsequent transformation of these partons 
into the observable hadrons. The latter, hadronization stage can be 
described by the standard string fragmentation framework 
\cite{AGIS}, followed by the decays of unstable primary hadrons, 
and is not further discussed here. The parton shower, 
multiple-interaction and beam-remnant aspects are discussed 
elsewhere~\cite{paper}.

\section{Comparisons with Data}
\label{migration}

In this section the model is compared with data. We will not make a detailed 
analysis of experimental results but use it to point out model dependences
and to constrain some model parameters. 

$2 \rightarrow 2$ parton interactions normally give rise to 2--jet events.
In leading--order QCD, the jets are balanced in transverse momenta in the 
centre of mass frame of the $\gast\p$ subsystem. Various effects,
such as primordial $k_{\perp}$, initial- and final-state bremsstrahlung,
tend to spoil this picture. This increases the $\d\sigma/\d p_{\perp}$ 
spectrum at any fixed $p_{\perp}$, since jets can be boosted up from lower 
$p_{\perp}$. In order to study jets above some 
$p_{\perp,\mathrm{min}}^\mathrm{jet}$, typically a
$p_{\perp,\mathrm{min}}^\mathrm{parton}=\frac{1}{2} 
 p_{\perp,\mathrm{min}}^\mathrm{jet}$ or less is required in the generation 
procedure.

\subsection{Inclusive $\e\p$ Jet Cross Sections}

\begin{figure} [!ht]
   \begin{center}  
     \mbox{\psfig{figure=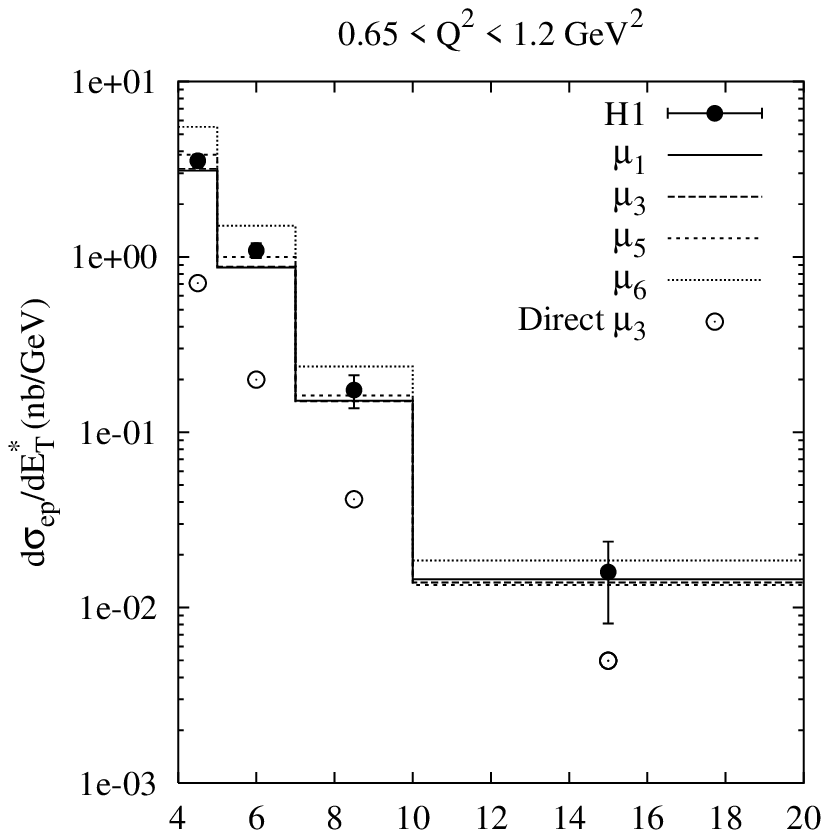,width=78mm}\hspace{-0.5cm}
	   \psfig{figure=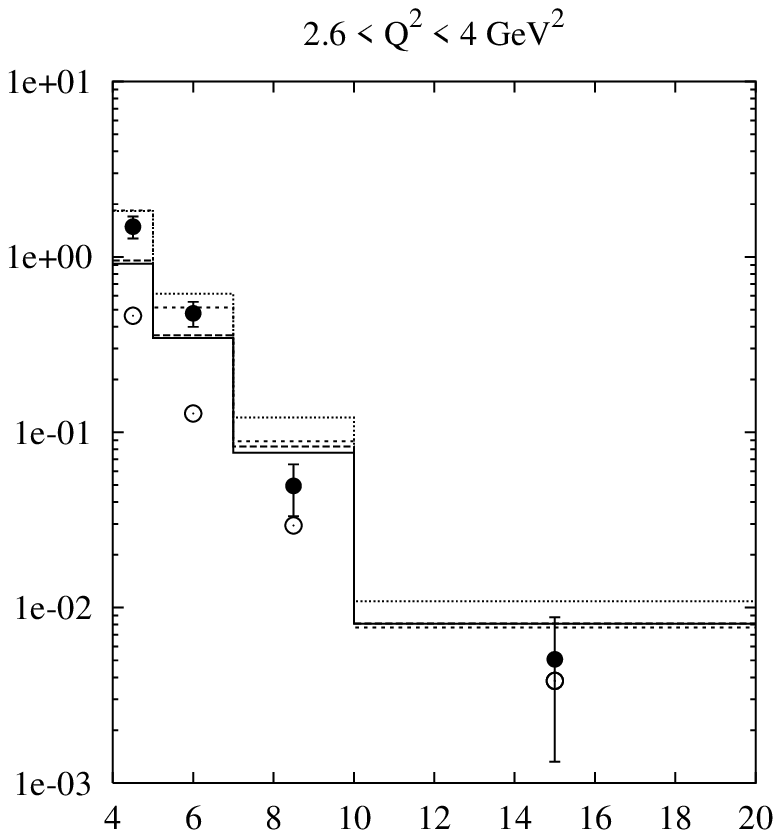,width=78mm}}
     \mbox{ }
     \mbox{\psfig{figure=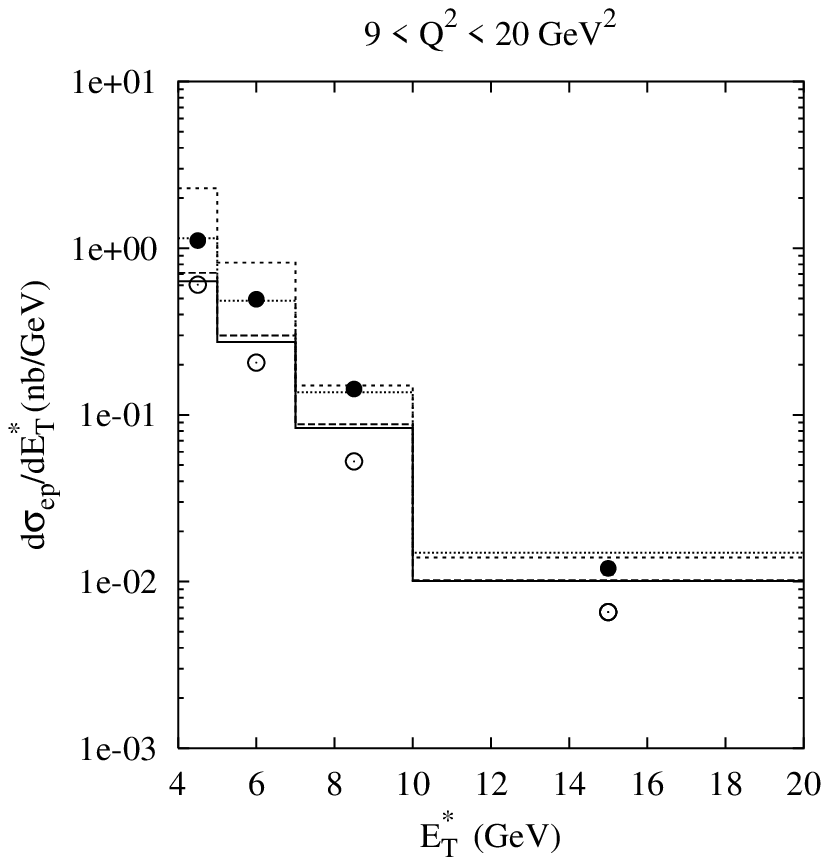,width=78mm}\hspace{-0.5cm}
	   \psfig{figure=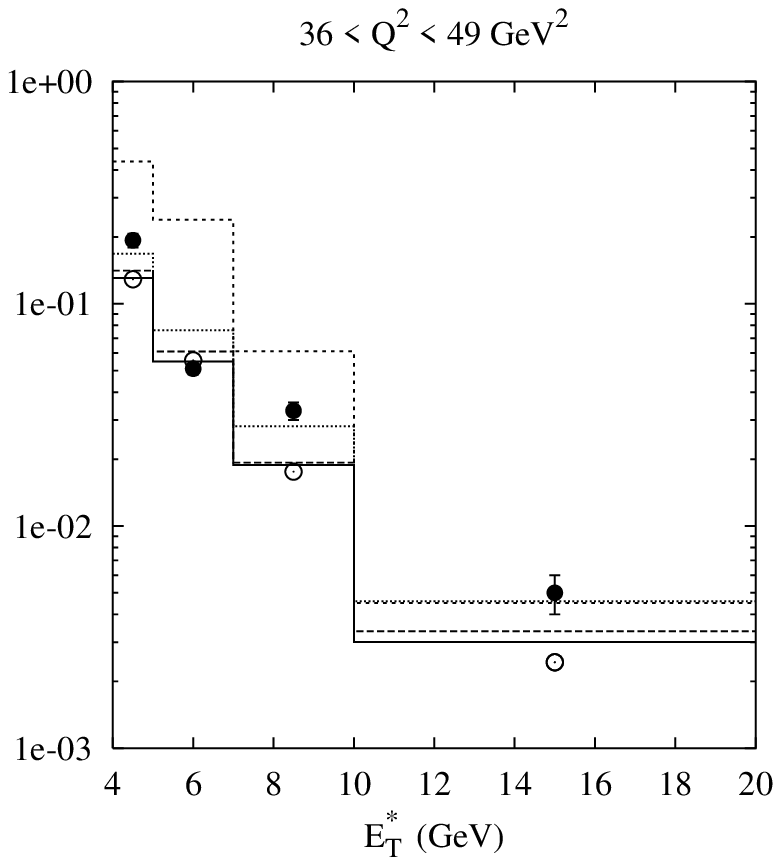,width=78mm}}
   \end{center}
\caption{{\it The differential jet cross section $\d\sigma_{\e\p}/\d E^*_{\perp}$ 
for jets with $-2.5<\eta^*<-0.5$ and $0.3<y<0.6$.
\label{fig:ET}}}
\end{figure}

\begin{figure} [!ht]
   \begin{center}       
     \mbox{\psfig{figure=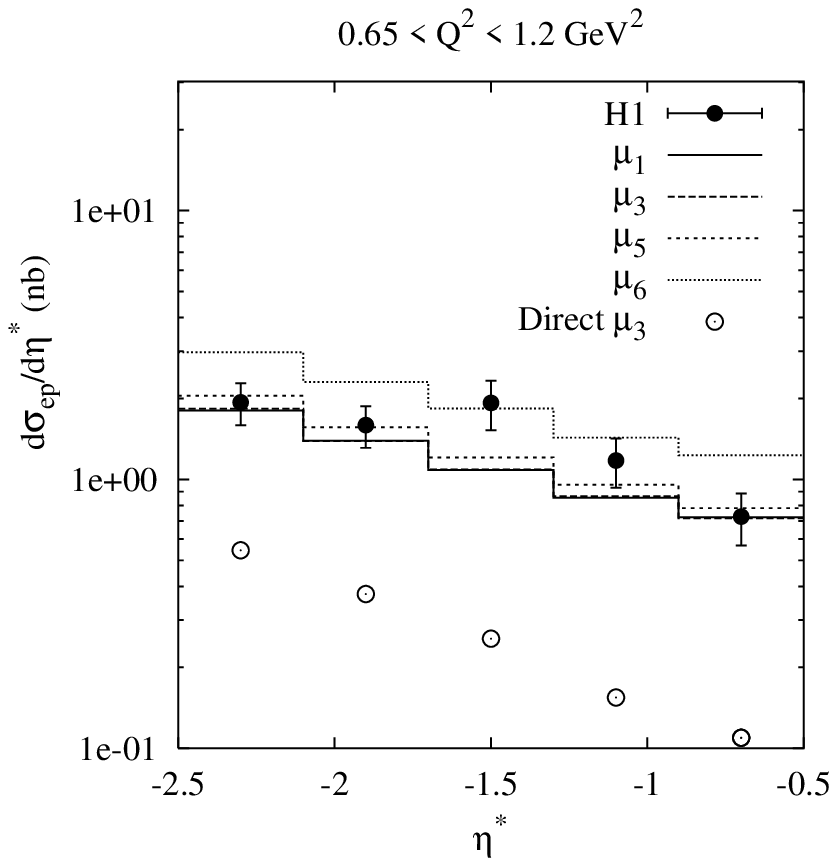,width=78mm}\hspace{-0.5cm}
	   \psfig{figure=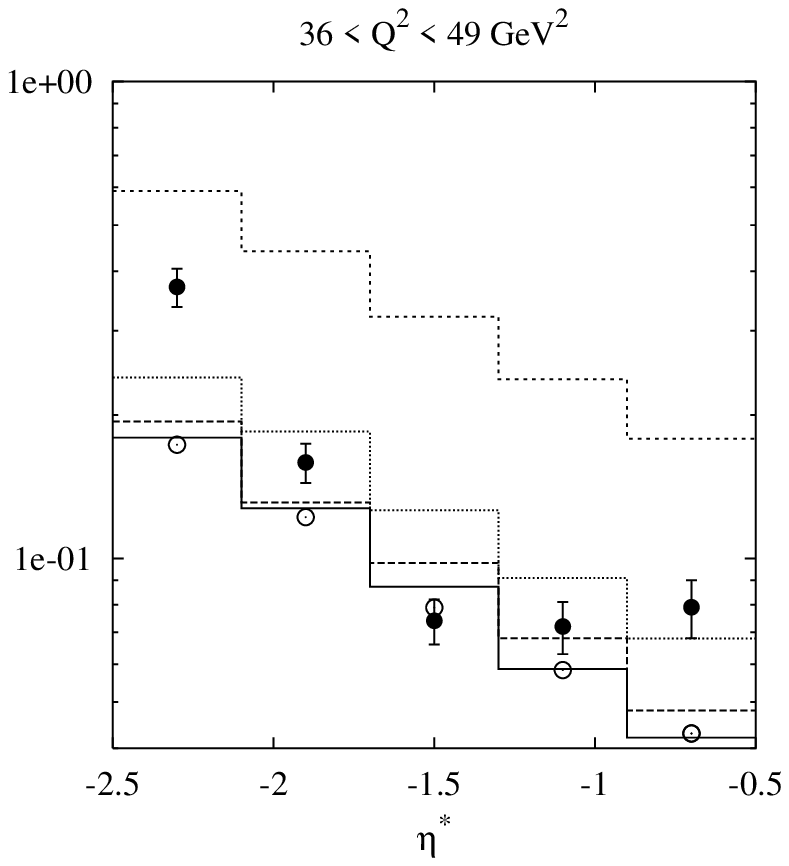,width=78mm}}
   \end{center}
\caption{{\it The differential jet cross section $\d\sigma_{\e\p}/\d \eta^*$ for 
jets with $E^*_{\perp}>5~\mathrm{GeV}$ and $0.3<y<0.6$.
\label{fig:eta}}}
\end{figure}

Inclusive $\e\p$ jet cross sections have been measured by the H1 collaboration 
\cite{lowQ2H1} in the kinematical range $0<Q^2<49~\mathrm{GeV}^2$ and 
$0.3<y<0.6$. The differential jet cross sections 
$\d\sigma_{\e\p}/\d E^*_{\perp}$ and $\d\sigma_{\e\p}/\d \eta^*$ in 
Fig.~\ref{fig:ET} and \ref{fig:eta} respectively, 
were produced with the HzTool~\cite{hz} package. The $E^*_{\perp}$ and 
$\eta^*$ are calculated in the $\gast\p$ centre of mass frame where the 
incident proton direction corresponds to positive $\eta^*$. 

For $\d\sigma_{\e\p}/\d E^*_{\perp}$ and $\d\sigma_{\e\p}/\d \eta^*$ 
data is available in nine different $Q^2$ bins; some of them are shown here
with similar results for the intermediate bins. The SaS~1D parton 
distribution together with a few different $\mu_i$ scales are used to model 
the resolved photon component. The other choices of scales, $\mu_2^2$ and 
$\mu_4^2$, interpolates between these results.

In the highest $Q^2$ bin the direct component is the dominant contribution; 
the virtuality of the photon is for most events of the order of or larger 
than the transverse momenta squared, $Q^2\gtrsim p_{\perp}^2$. However, the
resolved component is not negligible and all the scales $\mu_i$, except 
$\mu_1$, depend on the photon virtuality. This gives a larger resolved 
component in this region as compared to the the conventional choice, 
$\mu_1=p_{\perp}$. In the low $Q^2$ bin the $\mu_i$ scales do not differ 
much from $p_{\perp}$, i.e. the results are not sensitive to the scale 
choice $\mu_i$. The exception is $\mu_6$, which there overshoots the data. 
The $\mu_4=p_{\perp}^2+Q^2/2$ scale (not shown) gives nice agreement with 
data for all different $Q^2$ bins. 

Changing the photon parton distribution from SaS~1D to SaS~2D will give a 
slightly lower result for the low $Q^2$ bins (for most points within the 
size of the symbols). Using CTEQ~3L instead of GRV leading order as the 
proton parton distribution reduce the result in some $E_{\perp}^*$ and 
$\eta^*$ bins by half. The GRV higher order parton distribution give a 
slightly lower result (as compared to GRV leading order).

Since the VMD part dies out quickly with increasing photon virtuality,
multiple interactions will only be visible at low $Q^2$ (multiple 
interactions for the anomalous component is not in the model so far). 
The anomalous component dominates over the VMD component already at 
1~GeV$^2$. Therefore, multiple interactions for the VMD component can safely 
be neglected for the distributions shown in this section. 

\subsection{Forward Jets in $\e\p$}
\label{secfwdep}

Jet cross sections as a function of Bjorken-$x$, $x_{\mathrm{Bj}}$, for 
forward jet production (in the proton direction) have been measured at 
HERA~\cite{fwdjet}. The objective is to probe the dynamics of the QCD cascade 
at small $x_{\mathrm{Bj}}$. The forward jet is restricted in polar angle 
w.r.t. the proton and the transverse momenta $p_{\perp}^\mathrm{jet}$ should 
be of the same order as the virtuality of the photon, suppressing an 
evolution in transverse momenta. If the jet has a large energy fraction of 
the proton there will be a big difference in $x$ between the jet and the 
photon vertex; $x_{\mathrm{Bj}} \ll x_{\mathrm{jet}}$, allowing an evolution 
in $x$. The above restrictions will not eliminate the possibility of having a 
resolved photon, although the large $Q^2$ values are not in favour of it. 

\begin{figure} [!ht]
   \begin{center}     
   \mbox{\psfig{figure=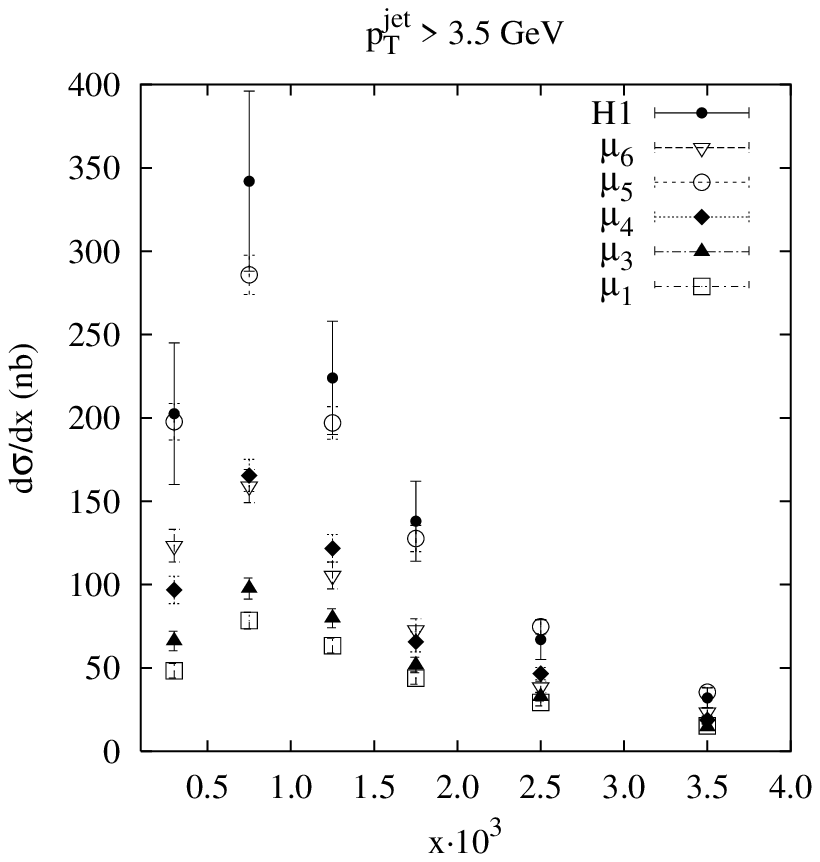,width=78mm}\hspace{-0.5cm}
	\psfig{figure=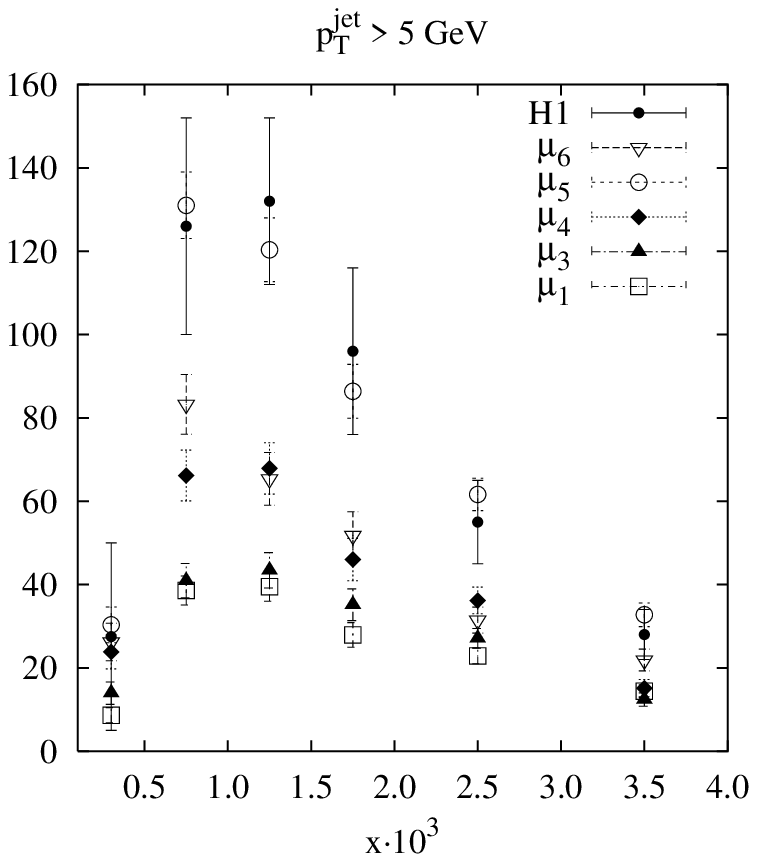,width=78mm}}  
   \end{center}
\caption{{\it Forward jet cross section as a function of $x$ compared with 
H1~data (with statistical and systematic uncertainties added in quadrature). 
Five different scales are shown at two 
different $p_{\perp}^\mathrm{jet}$ cuts, $3.5$ and $5~\mathrm{GeV}$. 
$x_\mathrm{jet}>0.035$, $0.5<(p_{\perp}^\mathrm{jet})^2/Q^2<2$ and 
$7^{\circ}< \theta_\mathrm{jet} < 20^{\circ}$. 
\label{fig:fwd}}}
\end{figure}

\begin{figure} [!ht]
   \begin{center}     
   \mbox{\psfig{figure=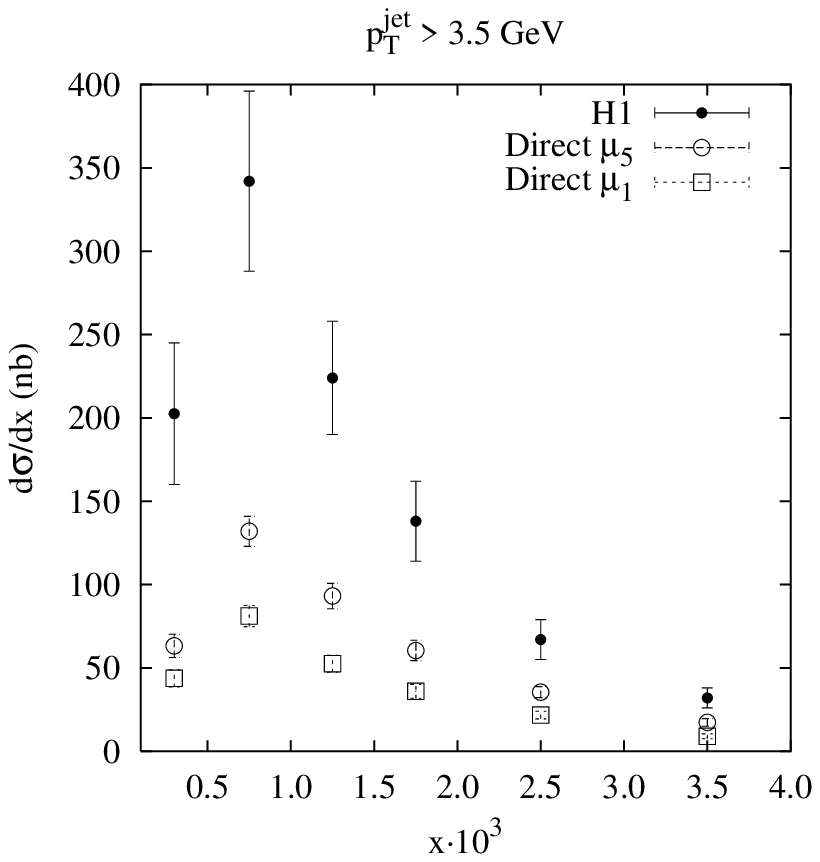,width=78mm}\hspace{-0.5cm}
	\psfig{figure=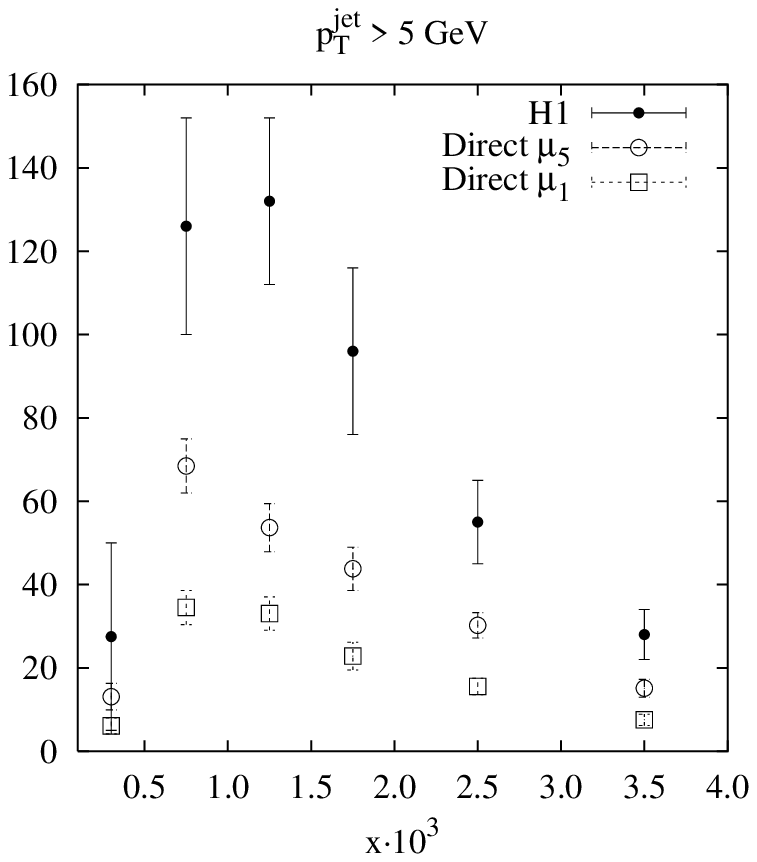,width=78mm}}  
   \end{center}
\caption{{\it Same as in Fig.~\ref{fig:fwd} but only the direct component is 
compared with data for two different scale choices.
\label{fig:fwd-dir}}}
\end{figure}

The HzTool routines~\cite{hz} were used to obtain the results in 
Fig.~\ref{fig:fwd}. Four different scales $\mu_i$ are shown. A larger forward 
jet cross section is obtained with a stronger $Q^2$ dependence, with the scale 
$\mu_5^2=p_{\perp}^2+Q^2$ in best agreement with data~\cite{LeifHannes}. The 
choice of scale does not only affect the resolved photon contribution but also 
the direct photon, arising from the scale dependence in the proton parton 
distribution, as seen in Fig.~\ref{fig:fwd-dir}. The rather large $Q^2$ values, 
$Q^2 \simeq (p_{\perp}^\mathrm{jet})^2$, suppresses VMD photons and favours the 
SaS~1D distribution which is the one used here, though the difference is small.

Note that the $\mu_6$ scale undershoots the forward jet cross section 
data and overshoots the inclusive jet distributions at low $Q^2$, 
so it is not a real alternative. As a further check, with more data 
accumulated and analysed, the $(p_{\perp}^\mathrm{jet})^2/Q^2$ interval 
could be split into several subranges which hopefully would help to 
discriminate between scale choices. 

\subsection{Importance of longitudinal resolved photons}

In this section we will study the importance of longitudinal resolved 
photons. A sensible $Q^2$--dependent scale choice, $\mu_3$, together with 
the SaS~1D distribution will be used throughout. 

\begin{figure} [!ht]
   \begin{center}    
     \mbox{\psfig{figure=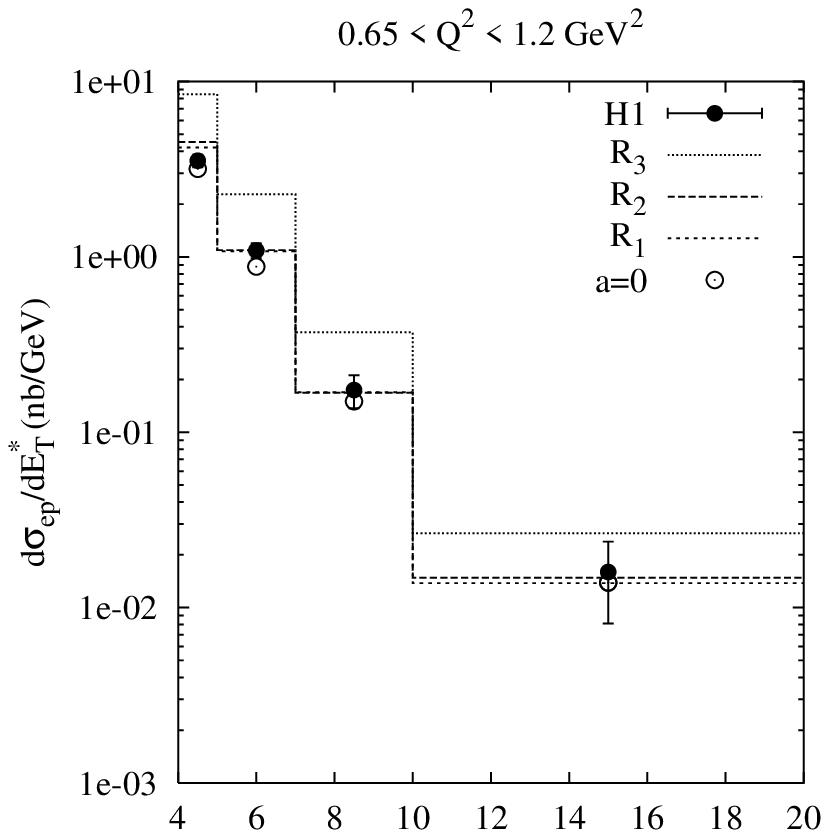,width=78mm}\hspace{-0.5cm}
	   \psfig{figure=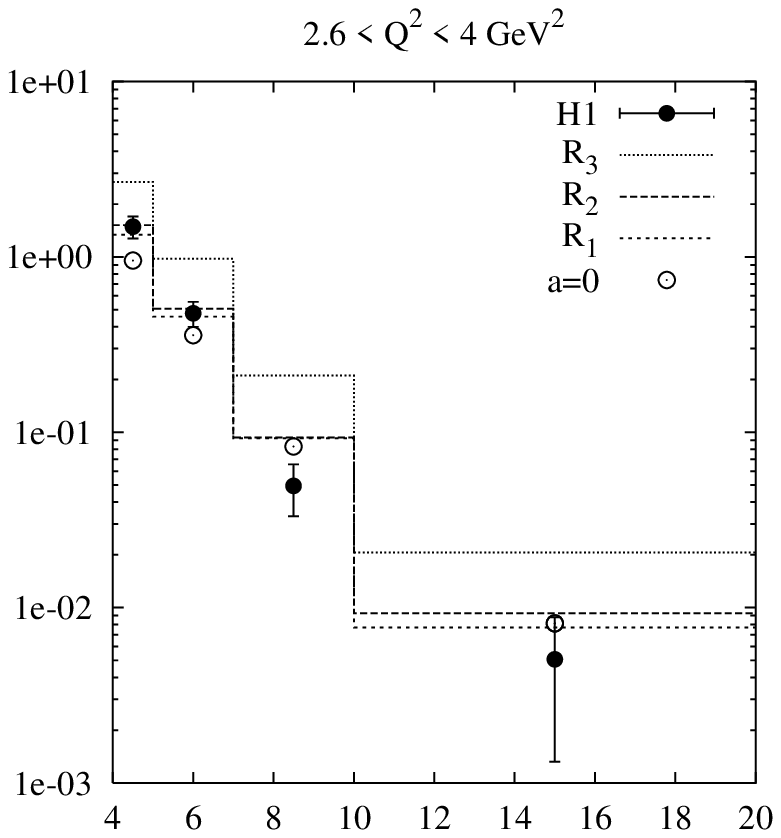,width=78mm}}
     \mbox{ }
     \mbox{\psfig{figure=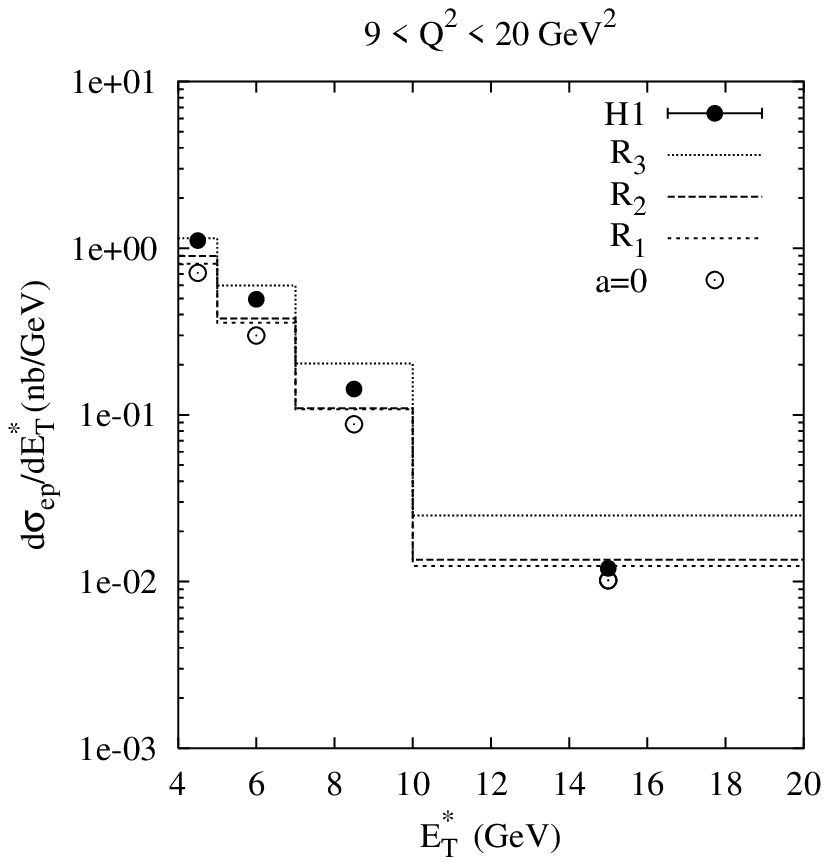,width=78mm}\hspace{-0.5cm}
	   \psfig{figure=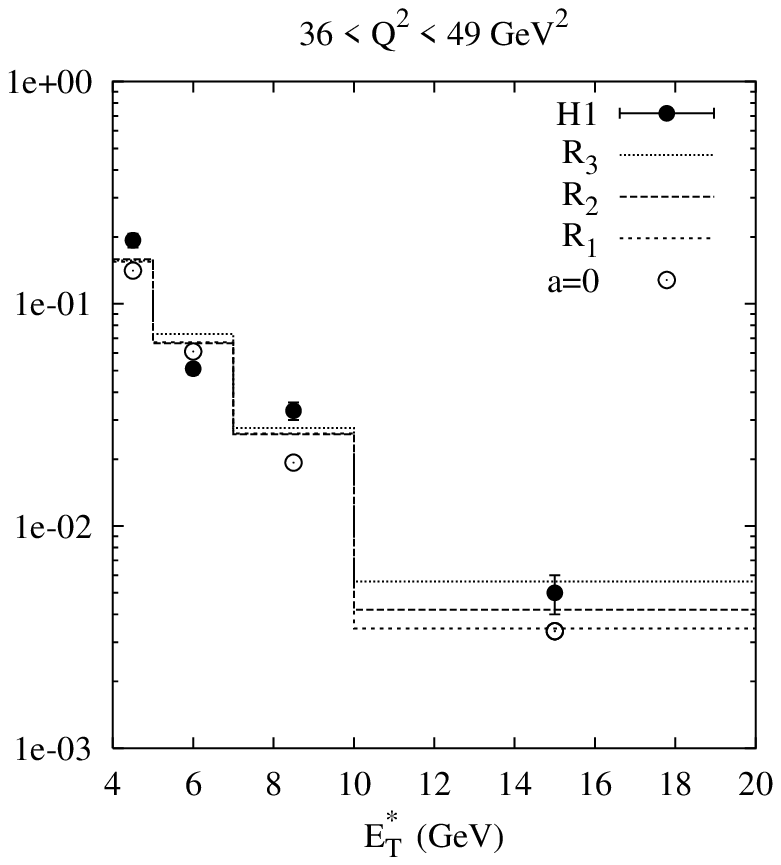,width=78mm}}
   \end{center}
\caption{{\it The differential jet cross section $\d\sigma_{\e\p}/\d E^*_{\perp}$ 
for jets with $-2.5<\eta^*<-0.5$ and $0.3<y<0.6$. 
\label{fig:ETL}}}
\end{figure}

With $a=1$ the different alternatives are shown in Fig.~\ref{fig:ETL} 
for the $\d\sigma_{\e \p}/\d E_{\perp}^*$ distributions together 
with the result from pure transverse photons, i.e. $a=0$. The importance 
of the resolved contributions decreases with increasing $Q^2$, see 
Fig.~\ref{fig:ET}, which makes the asymptotic behaviour less crucial. 
The onset of longitudinal photons governed by the $R_1$ and $R_2$ 
alternatives are favoured whereas the $R_3$ one overshoots data in the 
context of the other model choices made here. 

\begin{figure} [!ht]
   \begin{center}     
   \mbox{\psfig{figure=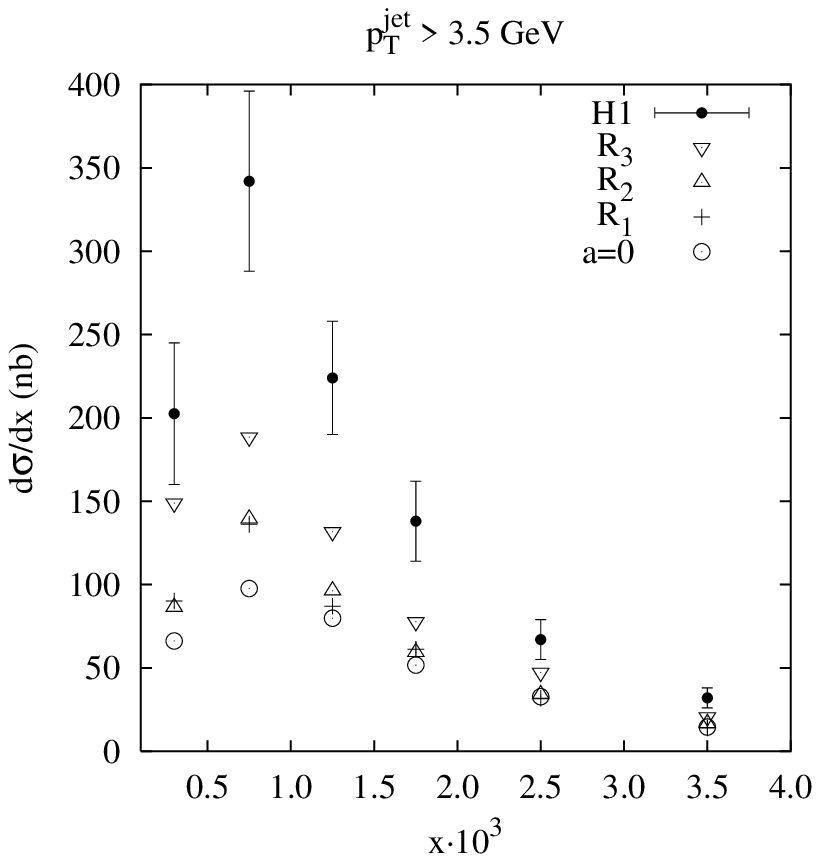,width=78mm}\hspace{-0.5cm}
	\psfig{figure=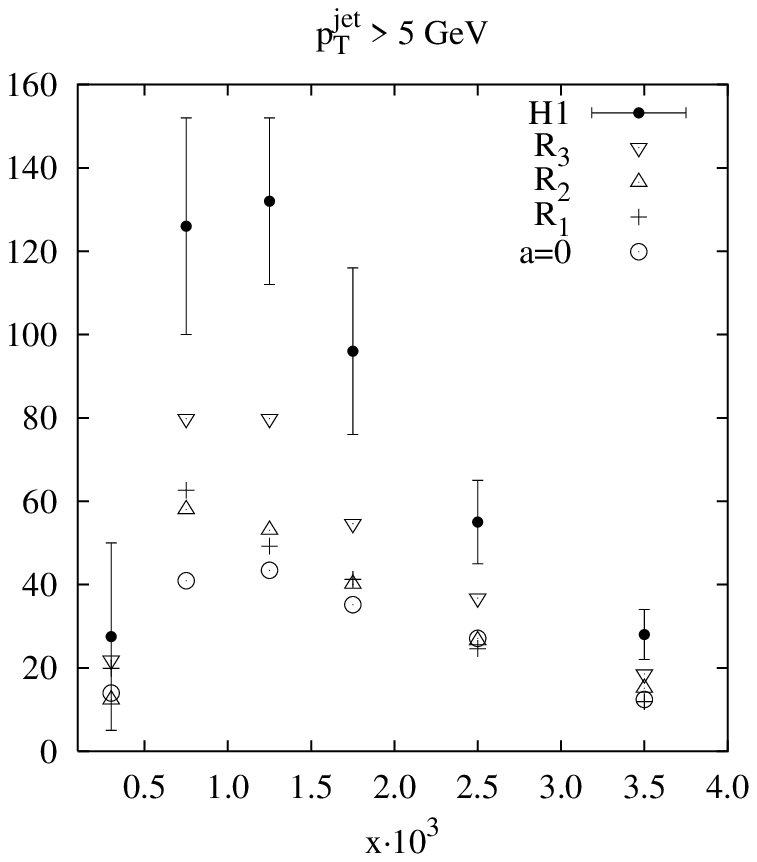,width=78mm}}  
   \end{center}
\caption{{\it Forward jet cross section as a function of $x$. The results
with three different alternatives of longitudinal resolved photons $R_i$ 
are compared with purely transverse ones, $a=0$, and data from H1.
\label{fig:fwdL}}}
\end{figure}

In Fig.~\ref{fig:fwdL} the same alternatives are shown for the forward jet 
cross sections. With this scale choice, $\mu_3$, none of the longitudinal 
resolved components (together with the direct contribution) are sufficient 
to describe the forward jet cross section. The resolved contribution with 
$R_3$ is about the same as the one obtained with the scale 
$\mu_5^2=p_{\perp}^2+Q^2$ (without longitudinal contribution); 
the difference in the total results originates from the difference in the
direct contributions, see Fig.~\ref{fig:fwd-dir}. With $R_1$ and $a=1$, the
$\mu_5$ scale (not shown) overshoots the data, but undershoots in combination 
with $\mu_4$.

The above study indicates, as expected, that longitudinal resolved photons
are important for detailed descriptions of various distributions. It cannot
by itself explain the forward jet cross section, but may give a significant 
contribution. Combined with other effects, for example, different scale 
choice, parton distributions and underlying events, it could give a 
reasonable description. The model(s) so far does not take into account the 
difference in $x$ distribution or the $k^2$ scale (of the 
$\gast \rightarrow \q\qbar$ fluctuations) between transverse and longitudinal 
photons. As long as the distributions under study allow a large interval in 
$x$ the average description may be reasonable. In a more sophisticated 
treatment these aspects have to be considered in more detail. 

\section{Summary and Outlook}

The field of photon physics is rapidly expanding, not least by the impact 
of new data from HERA and LEP. The prospects of building a Linear Collider, 
with its objective of high-precision measurements and to search for possible 
new physics, requires an accurate description of photon processes. 
The plan here is to have a complete description of the main physics aspects 
in $\gamma\p$ and $\gamma\gamma$ collisions, which will allow important cross 
checks to test universality of certain model assumptions. As a step forward, 
we have in this study concentrated on those that are of importance for the 
production of jets by virtual photons, and are absent in the real-photon case. 
While we believe in the basic machinery developed and presented here, we have 
to acknowledge the many unknowns --- scale choice, parton distribution sets, 
longitudinal contributions, underlying events, etc. --- that all give 
non--negligible effects. To make a detailed tuning of all these aspects was 
not the aim here, but rather to point out model dependences that arise from 
a virtual photon. 

When $Q^2$ is not small, naively only the direct component needs to be 
treated, but in practice a rather large contribution arises from resolved 
photons. For example, for high $Q^2$ studies like forward jet cross sections, 
Fig.~\ref{fig:fwd} and~\ref{fig:fwd-dir}, or inclusive differential jet cross 
sections, Fig.~\ref{fig:ET} and~\ref{fig:eta}. Resolved longitudinal 
photons are poorly understood and the model(s) presented here can be used to 
estimate their importance and get a reasonable global description. 
Longitudinal effects are in most cases small but of importance for fine--tuning. 

The forward jet cross section presented by H1~\cite{fwdjet} is well described
by an ordinary parton shower prescription including the possibility of having
resolved photons. The criteria that the $p_{\perp}^\mathrm{jet}$ should be of
the same order as $Q^2$, makes the scale choice crucial and in favour of data
is $\mu_5^2=p_{\perp}^2+Q^2$. With more data accumulated and analysed, the 
$(p_{\perp}^\mathrm{jet})^2/Q^2$ interval could be split into several subranges, 
which hopefully would help to discriminate between different scale choices.

After this study of jet production by virtual photons it is natural to connect 
it together with low--$p_{\perp}$ events. Clearly, a smooth transition from 
perturbative to non--perturbative physics is required. Further studies are 
needed and will be presented in a future publication. 

\subsubsection*{Acknowledgements}

We acknowledge helpful conversations with, among others, Jon Butterworth, 
Jiri Ch\'yla, Gerhard Schuler, Hannes Jung, Leif J\"onsson, Ralph Engel and 
Tancredi Carli.

\end{document}